\documentclass[fleqn,12pt,twoside]{article} 
\usepackage[headings]{espcrc1} 
  
\readRCS 
$Id: espcrc1.tex,v 1.2 2004/02/24 11:22:11 spepping Exp $ 
\usepackage{graphicx} 
\usepackage{epsfig} 
\usepackage[figuresright]{rotating} 
 
\newcommand{\AmS}{{\protect\the\textfont2 
  A\kern-.1667em\lower.5ex\hbox{M}\kern-.125emS}} 
 
\title{Cooling of Neutron Stars with Color Superconducting Quark Cores} 
 
\author{David Blaschke 
\address[GSI]{Theory Division, GSI mbH, D--64291 Darmstadt, Germany}% 
        \thanks{Bogoliubov Laboratory of Theoretical Physics, JINR, 
          141980, Dubna, Russia}, 
        Dmitri Voskresensky\addressmark[GSI]\thanks{Moscow Institute for  
Physics and Engineering, 115409 Moscow, Russia}
        and 
       Hovik Grigorian 
\address{Institut f\"ur Physik, Universit\"at Rostock, D-18051 Rostock, 
Germany} 
        \thanks{Department of Physics, Yerevan State University, 
          375025 Yerevan, Armenia} 
} 
        
\runtitle{Cooling of Neutron Stars} 
\runauthor{D. Blaschke} 
 
\begin{document} 
 
% typeset front matter 
\maketitle 
 
\begin{abstract} 
We show that within a recently developed nonlocal chiral quark model the  
critical density for a phase transition to color superconducting quark matter  
under neutron star conditions can be low enough for these phases to occur in  
compact star configurations with masses below 1.3 M$_\odot$. We study the  
cooling of these objects in isolation for different values of the  
gravitational mass and argue that, if the quark matter phase would allow  
unpaired quarks, the corresponding hybrid stars would cool too fast.  
The comparison with observational data puts tight constraints on possible  
color superconducting quark matter phases. Possible candidates with diquark  
gaps of the order of 10 keV - 1 MeV such as the "2SC+X" and the color spin  
locking (CSL) phase are presented.
\end{abstract} 
 
\section{Introduction} 
The cooling of compact stars is a complex problem which requires knowledge 
of the physics of strongly interacting matter and its coupling to leptonic 
degrees of freedom 
in a wide domain of temperatures, isospin asymmetries and densities from the 
solid-state like crust to the core of the star, where at supersaturation 
densities a transition to quark matter with a variety of possible 
superconducting phases is expected to occur.
We are witnessing a new era of compact star physics now when observational
data reach a level of accuracy which allows to discriminate between 
theoretical models.
One example is the recent mass measurement in a neutron star - white dwarf 
binary system where for the pulsar 
PSR 0751+1807 a mass of $2.1~\pm~0.2$ M$_\odot$ has been reported 
\cite{Nice:2005}. If such a mass value for a compact star would be settled
within the limits given by the presently reported 1 $\sigma$ level, this 
would rule out all hybrid star models with a quark matter core known up to now.
Another example is the upper limit for the surface temperature of the pulsar
PSR J0205+64 in the supernova remnant 3C58 \cite{Slane:2002} which is 
significantly below the {\it normal} cooling behavior, given the young age for 
this object associated with the historical supernova from AD 1181.
This implies a sensible dependence of the cooling processes on one of the 
characteristic parameters unknown for that compact star, such as the mass,
see Fig. 1.
For more observational constraints on neutron star properties, see \cite{NSC}
and Refs. therein.
It has been demonstrated recently in Ref. \cite{bgv2004}, that a satisfactory 
description of cooling data and structure of compact stars can be given
within a hadronic model without quark matter ({\it nuclear medium cooling} 
(NMC) scenario \cite{SVSWW,V01}), where medium effects are taken into account 
consistently.  
This approach could also satisfy the independent Log N - Log S test of a 
population synthesis model \cite{pgtb2004} and the brightness constraint 
\cite{G05}.
One particular result of the NMC approach was the formulation of a direct 
Urca (DU) constraint on the equation of state (EoS) under compact star 
conditions:
If a nuclear EoS allows the DU process to occur in typical 
compact star configurations with masses below $\sim ~1.5$ M$_\odot$,  
then this EoS has to be abandoned since otherwise a satisfactory description 
of cooling data cannot be obtained. 
For details, see also \cite{KV05,Klahn:2005}. 
The DU constraint holds in essence
also for stars with quark matter phases where, however, the pairing of all 
quark species with appropriate pairing gaps can result in an acceptable 
cooling phenomenology \cite{bkv,ppls,bgv,gbv}. 
In the present contribution we discuss 
the constraints on superconducting quark matter phases from the cooling 
phenomenology of stable hybrid stars when chiral quark matter models are used.

\section{Compact star cooling constraints on superconducting quark matter 
phases}

The state-of-the-art calculations for a three-flavor quark matter phase 
diagram within a chiral (NJL) quark model of quark matter and selfconsistently 
determined quark masses are described in Refs. 
\cite{Ruster:2005,Blaschke:2005,Abuki:2005}.
From these results follows that for the discussion of late cooling stages
when the temperature is well below the opacity temperature $T_{\rm opac}\sim 1$
MeV for neutrino untrapping four phases are relevant: the normal quark matter
(NQ), the two-flavor superconducting matter (2SC), a mixed phase of both
(NQ-2SC) and the color-flavor-locking phase (CFL).
The detailed structure of the phase diagram in these models still depends on 
the strength parameter $G_D$ of the diquark coupling (and of the formfactor 
of the momentum space regularization, see \cite{Aguilera:2004}).
For all values of  $G_D$ no stable hybrid stars with a CFL phase could be 
found yet, see \cite{Buballa:2004}, and Refs. therein.
We are left with the discussion of 2SC and NQ phases (the discussion of the 
NQ-2SC mixed phase brings no new aspects and will be omitted for brevity).

For the 2SC phase stable hybrid star configurations with masses even below 
$1.3$ M$_\odot$ have been obtained when a Gaussian formfactor regularization 
has been used. 
This phase has one unpaired color of quarks (say blue) for which the 
very effective quark DU process works and leads to a too fast cooling of the 
hybrid star in disagreement with the data \cite{gbv}.
We have suggested to assume a weak pairing channel which 
could lead to a small residual pairing of the hitherto unpaired blue quarks. 
We call the resulting gap $\Delta_X$ and show that for a 
density dependent ansatz
\begin{equation}
\label{gap}
\Delta_X(\mu)=\Delta_c~\exp[-\alpha(\mu-\mu_c)/\mu_c]~,
\end{equation}
with $\mu$ being the quark chemical potential, $\mu_c=330$ MeV,
$\alpha=10$ and $\Delta_c=1.0$ MeV an acceptable cooling 
phenomenology can be obtained \cite{gbv}, see Fig. 1.
The physical origin of the X-gap remains to be identified. It could occur, 
e.g., due to quantum fluctuations of color neutral quark sextett complexes
\cite{Barrois}. Such calculations have not yet been performed with the 
relativistic chiral quark models.

For sufficiently small $G_D$, the 2SC pairing may be inhibited at all.
In this case, due to the absence of this competing spin-0 phase with large 
gaps, one may invoke a spin-1 pairing channel in order to avoid the DU problem.
In particular the color-spin-locking (CSL) phase \cite{schafer}
may be in accordance with cooling phenomenology as all quark species are 
paired and the smallest gap channel may have a behavior similar to Eq. 
(\ref{gap}), see \cite{Aguilera:2005}. A consistent cooling calculation for 
this phase, however, requires the evaluation of 
neutrino emissivities and transport coefficients which is still to be 
performed.

Gapless superconducting phases can occur when the diquark coupling parameter
is small so that the pairing gap is of the order of the asymmetry in the 
chemical potentials of the quark species to be paired. Interesting 
implications for the cooling of gapless CFL quark matter have been conjectured
due to the particular behavior of the specific heat and 
neutrino emissivities  \cite{Alford:2005}. We find for reasonable values of 
$G_D$, however, that these phases do occur only at too high temperatures to be 
relevant for late cooling, if a stable hybrid configuration with these phases 
could be achieved at all.

\begin{figure}[ht] 
%\vspace{-0.5cm} 
\label{fig:1}
\parbox{0.8\textwidth}{ 
\epsfig{figure=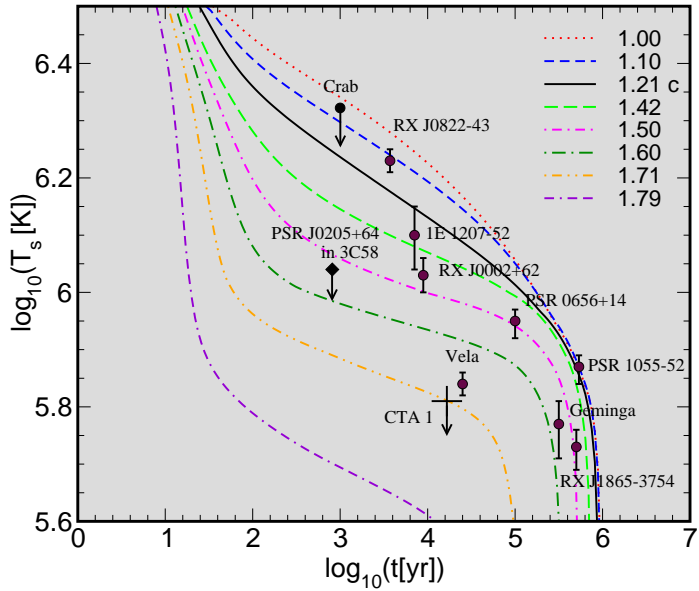,height=0.7\textwidth,angle=-90}} 
%\hfill
\hspace{-15mm}
\parbox{0.25\textwidth}{
\caption{Cooling curves for hybrid star configurations with Gaussian  
quark matter core in the 2SC+X phase with density-dependent X-gap. 
The labels correspond to the gravitational masses of the configurations  
in units of the solar mass.} } 
\end{figure} 

\section{Conclusions}

We have discussed that for modern phase diagrams for color superconducting 
three-flavor quark matter obtained within chiral quark models with 
selfconsistently determined quark masses out of the variety of possible
phases remain the 2SC+X and the CSL phases when the following conditions are 
applied:
%\begin{itemize}
- hybrid star stability against gravitational collapse,
- ompact star neutrality and $\beta$ equilibrium constraints,
- DU constraint for hadronic and quark matter.  
%\end{itemize}
However, the physical origin of the X-gap has not yet been clarified and 
a consistent cooling calculation of a hybrid star with CSL quark matter has 
still to be performed.
Gapless phases are unlikely to occur at low enough temperatures to be relevant 
for simulations of the late cooling of compact stars. 
\\[5mm] 
The present work has been supported in part by the Virtual Institute of the 
Helmholtz Association under grant No. VH-VI-041. 
H.G. acknowledges support by DFG grant No. 436 ARM 17/4/05, D.N.V. was 
supported by DFG grant No. 436 RUS 113/558/0-2 and RFBR grant NNIO-03-02-04008.


\begin{thebibliography}{9} 
\bibitem{Nice:2005} 
D.J. Nice et al., astro-ph/0508050.
\bibitem{Slane:2002}
P.O. Slane, D.J. Helfand, S.S. Murray, Astrophys. J. {\bf 571}, L45 (2002). 
\bibitem{bgv2004} 
D. Blaschke, H. Grigorian, and D.N. Voskresensky, 
%Cooling of Neutron Stars. Hadronic Model, 
%Preprint MPG-VT-UR 246/04 (2004). 
Astron. Astrophys. {\bf 424}, 979 (2004). 
%; [arXiv:astro-ph/0403170]. 
\bibitem{SVSWW}  
C. Schaab, D. Voskresensky, A.D. Sedrakian, F. Weber, and M.K. Weigel, 
Astron. Astrophys. {\bf 321}, 591 (1997).  
\bibitem{V01} 
 D.N.~Voskresensky, in book: "Physics of Neutron Star 
Interiors", 
Lecture Notes in Physics, Eds. D. Blaschke, N.K. Glendenning, A. Sedrakian, 
Springer, Heidelberg (2001), p. 467-502. 
\bibitem{pgtb2004} 
S. Popov, H. Grigorian, R. Turolla, and D. Blaschke, 
%{\it Population synthesis as a probe of neutron star thermal evolution}, 
%in preparation (2004). 
Astron. Astrophys. (2005) in press; [astro-ph/0411618]. 
\bibitem{G05}
H. Grigorian, astro-ph/0507052.
\bibitem{NSC}
J.M. Lattimer and M. Prakash, astro-ph/0405262.
\bibitem{KV05}
E.E. Kolomeitsev and D.N. Voskresensky, Nucl. Phys. {\bf A 759}, 373 (2005).
\bibitem{Klahn:2005}
T. Kl\"ahn et al., in preparation; H. Grigorian, D.N. Voskresensky, 
astro-ph/0507061.
\bibitem{bkv} 
D. Blaschke, T. Kl\"{a}hn, and D.N. Voskresensky, 
Astrophys. J. {\bf 533}, 406 (2000). 
\bibitem{ppls} 
D. Page, M. Prakash, J.M. Lattimer and A. Steiner, Phys. Rev. 
Lett. {\bf{85}} 2048 (2000). 
\bibitem{bgv} 
D. Blaschke, H. Grigorian and D.N. Voskresensky, 
Astron. \& Astrophys. {\bf 368}, 561 (2001). 
\bibitem{gbv}
H. Grigorian, D. Blaschke and D.N. Voskresensky,
Phys.\ Rev.\ C {\bf 71}, 045801 (2005).
\bibitem{Ruster:2005}
S.B. R\"uster, V. Werth, M. Buballa, I.A. Shovkovy and D.H. Rischke,
Phys. Rev. {\bf D 72}, 034004 (2005).
\bibitem{Blaschke:2005} 
D.Blaschke, S. Fredriksson, H. Grigorian, A.M. \"Oztas and F. Sandin,
Phys. Rev. {\bf D 72}, 065020 (2005).
\bibitem{Abuki:2005}
H. Abuki and T. Kunihiro, hep-ph/0509172.
\bibitem{Aguilera:2004}
D.N. Aguilera, D. Blaschke and H. Grigorian, 
Nucl. Phys. {\bf A 757}, 527 (2005). 
\bibitem{Buballa:2004}
M. Buballa, Phys. Rep. {\bf 407}, 207 (2005).
\bibitem{Barrois}
B. Barrois, Nucl. Phys. {\bf B 129}, 390 (1977).
\bibitem{schafer}
T. Sch\"afer, Phys. Rev. {\bf D 62}, 094007 (2000);\\
%\bibitem{schmitt}
TA. Schmitt, Q. Wang and D.H. Rischke, 
Phys. Rev. Lett. {\bf 91}, 242301 (2003).
\bibitem{Aguilera:2005}
D.N. Aguilera, D. Blaschke, M. Buballa, V.L. Yudichev, 
Phys. Rev. {\bf D 72}, 034008 (2005).
\bibitem{Alford:2005}
M. Alford, P. Jotwani, C. Kouvaris, J. Kundu and K. Rajagopal,
Phys. Rev. {\bf D 71}, 114011 (2005).
\end{thebibliography}
\end{document}